\documentclass[12pt]{article}
\usepackage{graphicx,psfrag}

\makeatletter

\newtoks\AIN@toks  \newtoks\AIN@stem
\newtoks\AIN@scrp  \newtoks\AIN@mode

\newif\ifAIN@group  \AIN@groupfalse

\newsavebox{\AIN@stemBox}  \newsavebox{\AIN@strut}

\def\tensor#1<#2>{\catcode`\ =9  \catcode`\^^M=9
   {\mathsurround=0pt  \everymath={\displaystyle}  \AIN@getStem{#2}
    \,\usebox{\AIN@strut}#1\!\usebox{\AIN@stemBox}
    \edef\@ct{\noexpand\AIN@step \the\AIN@scrp}\@ct  \usebox{\AIN@strut}}
  \futurelet\AIN@next\AIN@continue}

\def\AIN@continue{\catcode`\ =10  \catcode`\^^M=5  \let\AIN@step=\relax
  \ifx^\AIN@next  \,\let\AIN@step=\tensor       \fi
  \ifx_\AIN@next  \,\let\AIN@step=\tensor       \fi
  \ifx<\AIN@next  \,\let\AIN@step=\tensor       \fi
  \ifx`\AIN@next  \let\AIN@step=\AIN@interrupt  \fi  \AIN@step}

\def\AIN@interrupt#1#2'{\!#2\tensor}

\def\AIN@endTensor#1{\relax}

\def\AIN@putScrp#1#2>{\AIN@setMode{#1}
  \let\AIN@afterGet=\AIN@endScrp  \AIN@getToks #2>}

\def\AIN@endScrp{\usebox{\AIN@strut}\the\AIN@mode{\the\AIN@toks}
  \let\AIN@step=\AIN@putScrp
  \ifx>\AIN@next  \let\AIN@step=\AIN@endTensor  \fi  \AIN@step}

\def\AIN@setMode#1{\ifx#1^  \AIN@mode={^}  \else  \AIN@mode={^{}_}  \fi}

\def\AIN@getStem#1{\let\AIN@afterGet=\AIN@firstScrp  \AIN@getToks #1>\relax}

\def\AIN@firstScrp#1{\AIN@stem=\AIN@toks  \let\AIN@step=\AIN@getToks
  \ifx#1>  \AIN@mode={}  \AIN@toks={}  \def\AIN@step{\AIN@otherScrp>}
  \else    \AIN@setMode{#1}  \let\AIN@afterGet=\AIN@otherScrp  \fi  \AIN@step}

\def\AIN@otherScrp#1\relax{\AIN@scrp={#1}
  \ifx#1>  \let\AIN@step=\AIN@endTensor  \else  \let\AIN@step=\AIN@putScrp  \fi
  \sbox{\AIN@stemBox}{$\the\AIN@stem\the\AIN@mode{\the\AIN@toks}$}
  \edef\@ct{\noexpand\AIN@makeStrut \the\AIN@stem\relax}\@ct}

\def\AIN@makeStrut#1\relax{\def\densopt##1]##2{##2}  {\def\mathaccent##1{\relax}
  \def\dens##1{\let\noexpand\@st@p=\relax
    \ifx##1[  \let\noexpand\@st@p=\noexpand\densopt  \fi  \noexpand\@st@p##1}
  \let\sned=\dens
  \xdef\@ct{\noexpand\sbox{\AIN@strut}{$\noexpand\vphantom{#1}$}}}\@ct}

\def\AIN@getToks{\AIN@toks={}  \futurelet\AIN@next\AIN@nextTest}

\def\AIN@nextTest{\let\AIN@step=\AIN@addToks
  \ifx<\AIN@next        \let\AIN@step=\AIN@afterGet  \fi
  \ifx^\AIN@next        \let\AIN@step=\AIN@afterGet  \fi
  \ifx_\AIN@next        \let\AIN@step=\AIN@afterGet  \fi
  \ifx>\AIN@next        \let\AIN@step=\AIN@afterGet  \fi
  \ifx\:\AIN@next       \let\AIN@step=\AIN@conToks   \fi
  \ifx\bgroup\AIN@next  \AIN@grouptrue               \fi  \AIN@step}

\def\AIN@addToks#1{\ifAIN@group  \AIN@groupfalse
         \edef\@ct{\AIN@toks={\the\AIN@toks{\noexpand#1}}}
  \else  \edef\@ct{\AIN@toks={\the\AIN@toks\noexpand#1}}    \fi  \@ct
  \futurelet\AIN@next\AIN@nextTest}

\def\AIN@conToks#1#2{\edef\@ct{\AIN@toks={\the\AIN@toks\egroup#2\bgroup}}\@ct
  \futurelet\AIN@next\AIN@nextTest}

\def\@ApndToks#1#2{\edef\@act{\noexpand#1={\the#1#2}}\@act}
\def\@CopyToks#1#2{\edef\@act{\noexpand#2={\the#1}}\@act}

\newcommand{\dens}[2][1]{{
  \mathsurround=0pt  \everymath={\displaystyle}  \toks0={}
  \setbox0=\hbox{\lower 4.30554pt\hbox{$\mathchar"0365$}}  \dp0=0pt
  \count0=#1  \ifnum\count0 < 0  \multiply\count0 by -1  \fi
  \loop  \advance\count0 by -1  
    \@ApndToks{\toks0}{\copy0\crcr}
  \ifnum\count0 > 0
    \@ApndToks{\toks0}{\noalign{\kern -1pt\nointerlineskip}}
  \repeat
  \count0=#1  \skip3=0pt plus 3fil  \skip1=0pt plus 1fil
  \ifnum\count0 > 0
    \vbox{\ialign{\hskip\skip3##\hskip\skip1\crcr
      \the\toks0\noalign{\nointerlineskip}${#2}$\crcr}}
  \else
    \vtop{\ialign{\hskip\skip1##\hskip\skip3\crcr
      ${#2}$\crcr\noalign{\nointerlineskip\kern 2pt}\the\toks0}}
  \fi}\vphantom{#2}}
\newcommand{\sned}[2][1]{\dens[-#1]{#2}}

\newcommand{\Leftarrowfill}[0]{$\m@th  \mathord\Leftarrow  \mkern-6mu
  \cleaders\hbox{$\mkern-2mu \mathord= \mkern-2mu$}\hfill
  \mkern -6mu \mathord=$}

\newcommand{\pback}[2][1]{{
  \ifx2#1  \let\@rrow=\Leftarrowfill  \else  \let\@rrow=\leftarrowfill  \fi
  \mathchoice{\AIN@stemPullBack{#2}{\@rrow}}{\AIN@stemPullBack{#2}{\@rrow}}
             {\AIN@indxPullBack{#2}{\@rrow}}{\AIN@indxPullBack{#2}{\@rrow}}}}

\newcommand{\AIN@stemPullBack}[2]{{
  \vtop{\mathsurround=0pt  
  \ialign{##\crcr$\textstyle{#1}\strut$\crcr
    \noalign{\kern-0.4ex\nointerlineskip}{\tiny#2}\crcr}}
  \vphantom{\textstyle{#1}}}}

\newcommand{\AIN@indxPullBack}[2]{{
  \vtop{\mathsurround=0pt  
  \ialign{##\crcr\hfil$\scriptstyle{#1}$\hfil\crcr
    \noalign{\kern+0.4ex\nointerlineskip}{\tiny#2}\crcr}}}}

\newcommand{\address}[1]{\vbox{\let\\=\cr \normalsize \vskip 1em
  \lineskip\normallineskip \halign{\hfil##\hfil\crcr#1\crcr}}}

\newcommand{\topic}[1]{\medskip\noindent\textsl{#1}\medskip\noindent}

\newcommand{\mbb}[1]{\mathrm{#1}}
\newcommand{\mfs}[1]{\mathcal{#1}}

\DeclareFontFamily{U}{bbm}{}          
\DeclareFontShape{U}{bbm}{m}{n}{<5><6><7><8><9><10> gen * bbm 
  <10.95> bbm10 <12><14.4> bbm12 <17.28><20.74><24.88> bbm17}{}
\DeclareMathAlphabet{\mathbb}{U}{bbm}{m}{n}
\renewcommand{\mbb}[1]{\mathbb{#1}}
\DeclareFontFamily{U}{rsfs}{}         
\DeclareFontShape{U}{rsfs}{m}{n}{<5> rsfs5 <6><7> rsfs7 
  <8><9><10><10.95><12><14.4><17.28><20.74><24.88> rsfs10}{}
\DeclareMathAlphabet{\mathfs}{U}{rsfs}{m}{n}
\renewcommand{\mfs}[1]{\mathfs {#1}}

\newenvironment{eqtableau}[2][0pt]{\@begineqtableau{#1}{#2}}{\@endeqtableau}

\def\@begineqtableau#1#2{
  \vcenter\bgroup\openup1\jot
    \mathsurround=0pt  \everymath={\displaystyle}
    \dimen0=#1  \count0=#2  \toks0={\strut}  \toks1={##}
		\def\\{\crcr\noalign{\vskip\dimen0}}
    \ifnum\count0 > 0
      \loop  \advance\count0 by -1
        \@ApndToks{\toks0}{$\hfil\the\toks1$&${}\the\toks1\hfil$}
      \ifnum\count0 > 0
        \@ApndToks{\toks0}{&}
      \repeat
    \else  \@ApndToks{\toks0}{\hfil$\the\toks1$\hfil}  \fi
    \edef\@act{\noexpand\ialign\bgroup\the\toks0\noexpand\crcr}  \@act}
\def\@endeqtableau{\crcr\egroup\egroup}

\def\@begineqset{\begingroup  \mathsurround=0pt  \let\\=\@eqsetcr%
  \stepcounter{equation}\def\@currentlabel{\p@equation\theequation}%
  $$\everycr={}  \everymath={\displaystyle}  \openup\jot
    \tabskip=\@centering  \halign to\displaywidth\bgroup
    \hfil$##$\tabskip=0pt&${}##$\hfil\tabskip=\@centering&\relax
      \llap{##}\tabskip=0pt\crcr}
\def\@endeqset{\egroup\global\advance\c@equation by -1$$
  \endgroup\@ignoretrue}

\def\@eqsetcr{&\if@eqnsw\@eqnnum\stepcounter{equation}\fi
  \global\@eqnswtrue\crcr}

\newcommand{\grad}[0]{\nabla}
\newcommand{\implies}[0]{\Rightarrow}
\newcommand{\Lie}[0]{\mfs{L}}
\newcommand{\Tr}[0]{\mathrm{Tr}}
\newcommand{\tsfrac}[2]{{\textstyle\frac{#1}{#2}}}
\newcommand{\Com}[0]{\mbb{C}}
\renewcommand{\Re}[0]{\mbb{R}}

\makeatother

\newcommand{\four}[1]{\tensor^4<#1>\relax}
\newcommand{\emA}{\mbb{A}}
\newcommand{\emE}{\mbb{E}}
\newcommand{\scri}{\mfs{I}}
\newcommand{\ERad}{\ensuremath E_\infty^\mathrm{Rad}}
\newcommand{\area}{\ensuremath A_\Delta^{}}

\renewcommand{\theequation}{\arabic{equation}}

\begin{document}

\title  {Isolated Horizons:\\ A Generalization of Black Hole Mechanics}
\author {Abhay Ashtekar, Christopher Beetle and Stephen Fairhurst\\
  \address{Center for Gravitational Physics and Geometry\\
           Department of Physics, The Pennsylvania State University\\
           University Park, PA 16802}}
\date   {December 17, 1998}
\maketitle

\begin{abstract}
A set of boundary conditions defining a \textit{non-rotating isolated horizon} 
are given in Einstein-Maxwell theory.  A space-time representing a black hole 
which itself is in equilibrium but whose exterior contains radiation admits 
such a horizon .  Physically motivated, (quasi-)local definitions of the mass 
and surface gravity of an isolated horizon are introduced.  Although these 
definitions do not refer to infinity, the quantities assume their standard 
values in Reissner-Nordstr\"om solutions.  Finally, using these definitions, 
the zeroth and first laws of black hole mechanics are established for isolated 
horizons.
\end{abstract}

\begin{figure}
  \begin{center}
    \begin{minipage}{1.5in}
      \begin{center}
				\psfrag{Ifs}{\Large $\scri^+$}
				\psfrag{Mcal}{\Large $\mathcal{M}$}
				\psfrag{i+}{\Large $i^+$}
				\psfrag{i0}{\Large $i^0$}
				\psfrag{Delta}{\Large $\Delta$}
				\psfrag{Mpln}{\Large $M$}
        \resizebox{!}{4cm}{\includegraphics{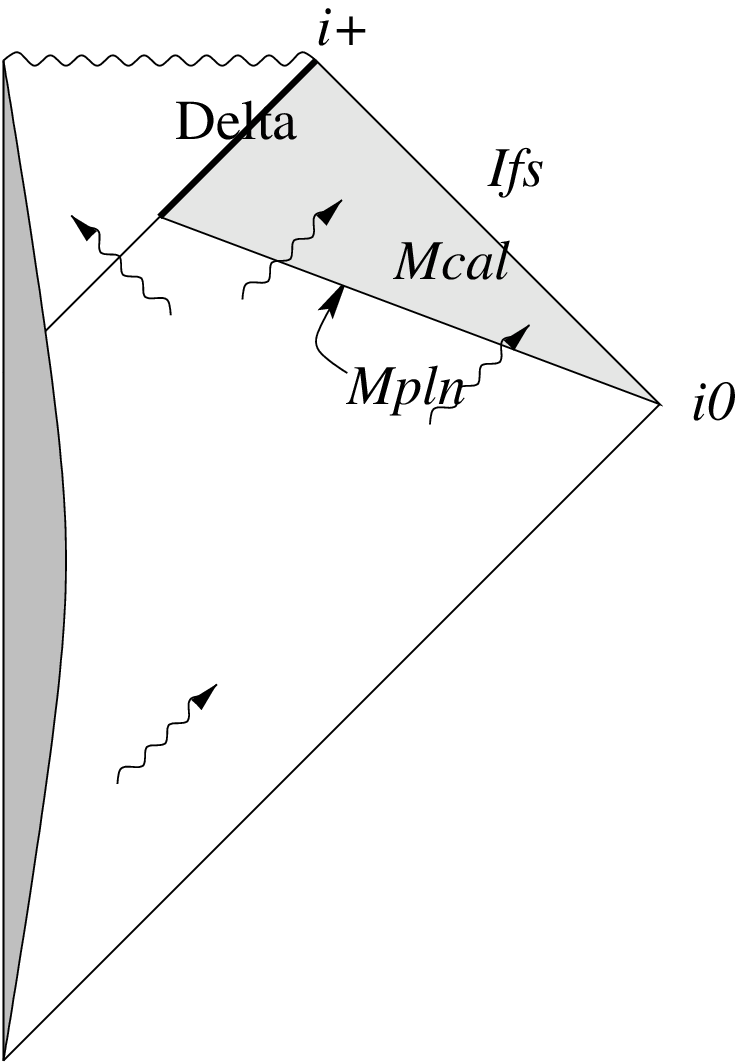}}\\(a)
      \end{center}
    \end{minipage}
    \hspace{.5in}
    \begin{minipage}{4in}
      \begin{center}
        \includegraphics[height=4cm]{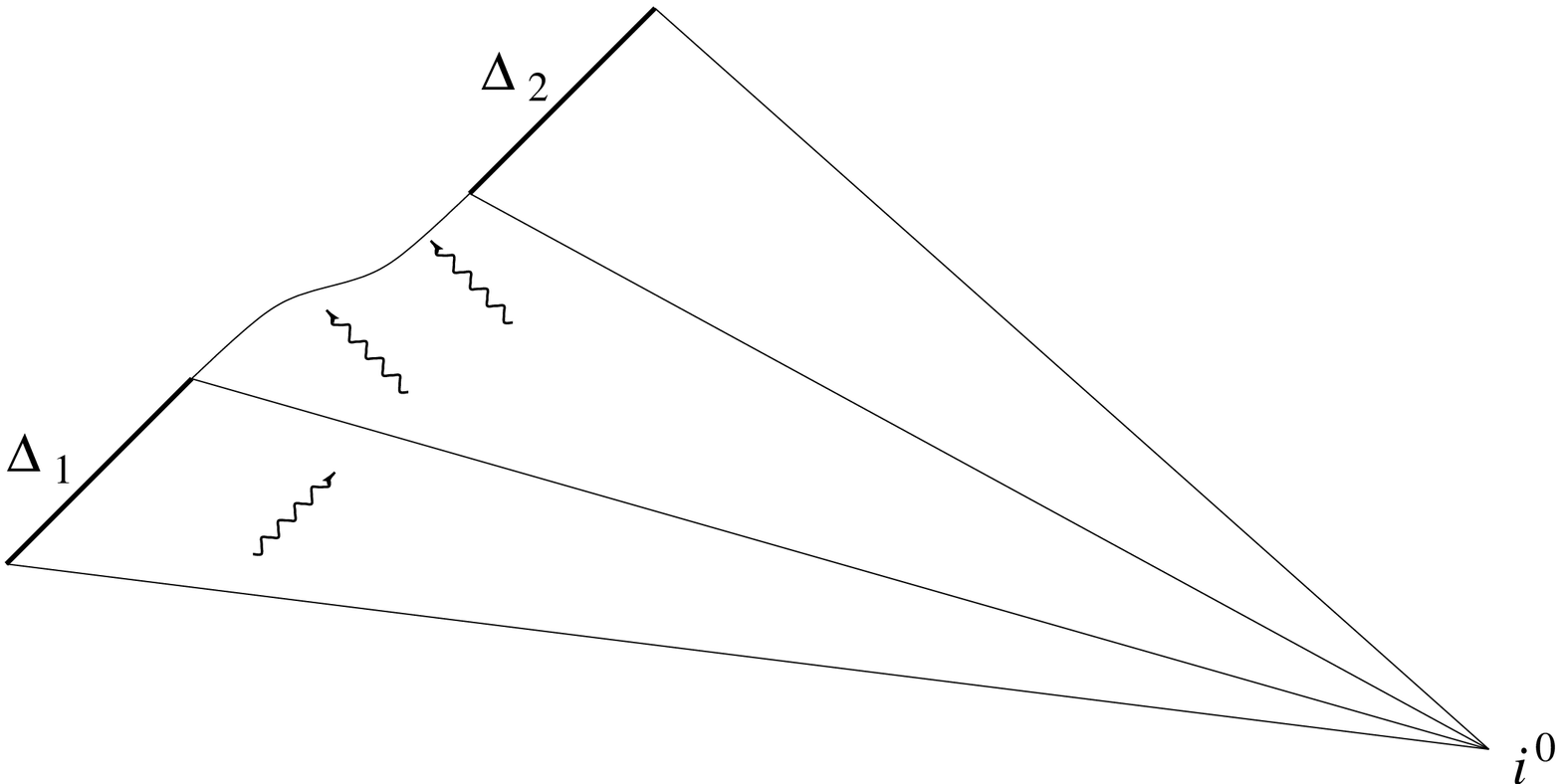}\\(b)
      \end{center}
    \end{minipage}
    \caption{(a)\quad A typical gravitational collapse.  The portion $\Delta$ 
      of the horizon at late times is isolated.  The space-time $\mathcal{M}$ 
      of interest is the triangular region bounded by $\Delta$, $\scri^+$ and 
      a partial Cauchy slice $M$.  \quad(b)\quad Space-time diagram of a black 
      hole which is initially in equilibrium, absorbs a small amount of 
      radiation, and again settles down to equilibrium.  Portions $\Delta_1$ 
      and $\Delta_2$ of the horizon are isolated.}\label{exam}
  \end{center}
\end{figure}

The similarity between the laws of black hole mechanics and those of ordinary 
thermodynamics is one of the most remarkable results to emerge from classical 
general relativity \cite{jb,bch,red}.  However, in the standard formulation, 
the zeroth and first laws apply only to stationary black holes.  From a 
physical perspective, the requirement of stationarity seems too strong since 
it excludes situations in which there may be radiation far away from the black 
hole.  Indeed, in a realistic gravitational collapse (depicted in figure 
\ref{exam}a), although one expects the horizon to reach an equilibrium state 
at late times, one also expects gravitational and other radiation to be 
present near null infinity.  One would hope that the familiar laws of black 
hole mechanics continue to hold in such situations.  The purpose of this 
letter is to show that this expectation is correct: We will outline a new 
framework to describe a generic (non-rotating) isolated black hole and extend 
the laws of black hole mechanics to this broader class of space-times.  This 
framework also serves as the point of departure for analyzing the quantum 
geometry of horizons non-perturbatively, which in turn provides a statistical 
mechanical explanation of the Bekenstein-Hawking entropy associated with the 
horizon \cite{abck}.

The key idea is to replace the notion of a stationary black hole with that of 
an {\it isolated horizon}.  This can be identified with a portion of the event 
horizon which is in equilibrium, i.e., across which there is no flux of 
gravitational radiation or matter fields.  Examples of isolated horizons are 
shown in figures \ref{exam}a and \ref{exam}b.  For technical simplicity, in 
this letter we will restrict our attention to non-rotating horizons.

Initially, the difference between static black holes and non-rotating isolated 
horizons may appear rather small.  Therefore, one might expect the required 
extension of the laws of black hole mechanics to be straightforward.  However, 
this is not the case.  Technically, the generalization involved is enormous: 
In Einstein-Maxwell theory, while there is only a three-parameter family of 
static black hole solutions, the space of solutions with isolated, 
non-rotating horizons is {\it infinite} dimensional \cite{abf2}.  The 
conceptual non-triviality of the problem runs even deeper and is brought out 
by the following considerations.  To formulate the zeroth and first laws, one 
needs a definition of the `extrinsic parameters' of the black hole --- 
particularly the surface gravity $\kappa$ and the mass $M$.  In the static 
context, $M$ is taken to be the ADM mass and $\kappa$ is taken to be the 
acceleration of that static Killing vector field which is unit at infinity.  
Although these parameters are associated with the black hole, they cannot be 
constructed from the space-time geometry only near the horizon; they are 
genuinely global concepts.  Now, if one has a non-static isolated horizon, the 
ADM mass \textit{cannot} be identified with the mass of the black hole since 
it also includes the mass associated with the radiative fields far away from 
the horizon.  Similarly, in the absence of a global Killing field, the 
definition of $\kappa$ is far from obvious: the above prescription for 
normalizing the null generator of the horizon, whose acceleration defines 
$\kappa$, is no longer applicable.  Therefore even the formulation --- let 
alone the proof --- of the zeroth and first laws appears problematic at first.  
However, we will show that these problems can be overcome.

The calculations reported in this letter are carried out in the 
`connection-dynamics' framework for general relativity \cite{blue}.  The basic 
gravitational degrees of freedom in this framework consist of an SL$(2,\Com)$ 
soldering form $\tensor<\sigma_a^AA'>$ and a self-dual SL$(2,\Com)$ connection 
$\tensor^4<A_a_A^B>$ over space-time.  (Here, the lower case latin letters are 
used for space-time indices and the upper case primed and unprimed indices 
refer to SL$(2,\Com)$ spinors.)  The electromagnetic U$(1)$ connection will be 
denoted by $\four\emA_a$.

\topic{Boundary Conditions}

A \textit{non-rotating, isolated horizon} is a 3-dimensional hypersurface 
$\Delta$ in space-time which satisfies the following conditions:

\begin{enumerate}

\item $\Delta$ is null, topologically $S^2\times \Re$, and equipped with a 
preferred foliation by 2-surfaces $S_\Delta$ transverse to its null normal 
$\ell^a$.  We will denote the other null normal to the foliation by $n^a$ and 
partially fix the normalization by setting $\ell^a n_a = -1$ and requiring the 
pull-back to $\Delta$ of $n_a$ to be curl-free.

\item All equations of motion are satisfied at $\Delta$.

\item The surface $\Delta$ is isolated in the sense that there is no flux of 
radiation either through or along it: $\tensor <R_\pback[2]{a}b> <\ell^b>=0$ 
and $\tensor <R_\pback[2]{a}b> <n^b>=0$, where $\Leftarrow$ denotes 
the pull-back to the foliation two-spheres $S_\Delta$.

\item $\Delta$ is a non-rotating, non-expanding future boundary of the 
space-time $\mathcal{M}$ under consideration: $\ell^a$ is geodesic and 
expansion-free and $n^a$ is shear-free with negative expansion $\theta_{(n)}$
which is constant on each leaf of the preferred foliation%
\footnote{This last condition guarantees the uniqueness of the preferred 
foliation of $\Delta$.  For a more detailed statement of the boundary 
conditions, see \cite{abf2}.}.

\end{enumerate}

Note that these conditions are \textit{local} since they are enforced 
\textit{only} at points of $\Delta$ and well-defined despite the residual 
rescaling freedom $(\ell^a, n_a) \rightarrow (F\ell^a, F^{-1}n_a)$ with $F$ 
constant on each $S_\Delta$.  Let us summarize the consequences of these 
conditions which are directly relevant here.  (For a more complete discussion, 
see \cite{abf2}.)  First, the pull-backs to $\Delta$ of the Maxwell field, its 
dual, and the space-time metric are spherically symmetric.  However, the 
space-time geometry need not be spherically symmetric even in a neighborhood 
of $\Delta$; the situation is rather similar to that at null infinity.  
Second, the area $\area$ of the spherical sections $S_\Delta$ and the electric 
and magnetic charges, $Q$ and $P$, contained within them are constant in time%
\footnote{Note, however, that these are not fixed constants and may take
different values in different space-times.}.  %
Furthermore, the pull-back $g_{\pback{ab}}$ to $\Delta$ of the metric 
satisfies $\Lie_\ell \, g_{\pback{ab}} = 0$, but $\ell^a$ need not be a 
Killing vector of the full metric $g_{ab}$ even at the horizon.  Third, it is 
easy to show that $\ell^a$ is twist- and shear-free and $n^a$ is twist-free.  
Finally, the pull-back $\tensor ^4<F_\pback[2]{ab}^{AB}>$ to $S_\Delta$ of the 
curvature of $\tensor ^4<A_a_A^B>$ satisfies
\begin{equation}\label{curv} 
  \tensor ^4<F_\pback[2]{ab}^{AB}> = -2i (\Phi_{11} - \Psi_2) \,
	  \tensor ^2 <\epsilon_{ab}> <\iota^{(A}> <o^{B)}>,
\end{equation}
where $\Psi_2$ and $\Phi_{11}$ are Newman-Penrose components of the Weyl and 
Ricci tensors, $\tensor ^2<\epsilon>$ is the volume form on the foliation two 
spheres, and the spinors $\iota^A$ and $o^A$ satisfy%
\footnote{Note that the metric $g_{ab}$ has signature $(-,+,+,+)$.} %
$i \tensor <\sigma^a_AA'> <\iota^A> <\bar\iota^A'> = n^a$ and $i \tensor 
<\sigma^a_AA'> <o^A> <\bar o^A'> = \ell^a$.

\topic{The Zeroth Law}

To define the surface gravity $\kappa$ of an isolated horizon, we must first 
fix the normalization of $\ell^a$; $\kappa$ is then the acceleration of the 
`correctly' normalized $\ell^a$.  Recall that $\ell^a$ itself is free of 
shear, expansion and twist, whence we cannot fix its normalization through its 
intrinsic properties at the horizon.  However, since the expansion 
$\theta_{(n)}$ of $n^a$ is strictly negative, we can always fix the 
normalization of $n^a$ by setting $\theta_{(n)}$ to any given value.  This in 
turn fixes the normalization of $\ell^a$ since $\ell^a n_a = -1$ and exhausts 
the residual rescaling freedom completely.  Furthermore, if in the 
Reissner-Nordstr\"om solutions $\ell^a$ is taken to be the restriction to 
$\Delta$ of the properly normalized static Killing field, then $\theta_{(n)} = 
-2/r_\Delta$, where $r_\Delta$ is the radius of $\Delta$ (i.e., $\area = 4\pi 
r_\Delta^2$).  Since we want to include this family in our analysis, let us 
require $n^a$ (and hence $\ell^a$) \textit{always} be normalized so 
$\theta_{(n)} = -2/r_\Delta$.  We now \textit{define} the surface gravity 
$\kappa$ of an isolated horizon by
\begin{equation}
\ell^b \grad_b \ell^a =: \kappa \ell^a.
\end{equation}
By construction, this expression reproduces the usual surface gravity in 
Reissner-Nordstr\"om space-times.  Furthermore, in the Einstein-Maxwell case, 
one can now show that for all space-times satisfying our boundary conditions, 
$\kappa$ is given by:
\begin{equation}\label{kappa}
  \kappa = -r_{\!\Delta} \, \Psi_2
\end{equation}
Finally, note that $\kappa$ is entirely local to the surface $\Delta$.

With this definition of surface gravity, the zeroth law of black hole 
mechanics can now be derived for a generic isolated horizon via a simple 
topological argument.  The key point is that the pull-back to any $S_\Delta$ 
of the self-dual connection $\tensor ^4<A_a_A^B>$ is a connection on the spin 
bundle of that 2-sphere.  Since the Chern number of the spin bundle over $S^2$ 
is unity, (\ref{curv}) gives
\begin{equation}\label{zero}
  2\pi i = \oint_{S_\Delta} \tensor ^4<F^AB> <\iota_A> <o_B> = i 
    (\Phi_{11} - \Psi_2) \area  \quad\implies\quad
  \kappa = -r_\Delta \left( \Phi_{11} - \tsfrac{2\pi}{\area} \right).
\end{equation}
In the Einstein-Maxwell case considered in this letter, the value of 
$\Phi_{11}$ at $\Delta$ is completely determined in terms of $Q$, $P$ and 
$\area$.  Since these are all constant on $\Delta$, $\kappa$ must also be 
constant on the horizon.  This result is precisely the zeroth law of black 
hole mechanics.

\topic{The First Law}

To derive the first law, we must first definite the mass of an isolated 
horizon.  Our definition will be motivated by the Hamiltonian framework.

A key question faced by any set of boundary conditions is whether they enable 
one to formulate an action principle.  For boundary conditions at infinity, it 
is well-known that the answer is in the affirmative: In the absence of 
internal boundaries, an action principle can be found by adding a boundary 
term at infinity to the standard bulk action \cite{blue}.  Similarly with our 
boundary conditions, one can again obtain a well-defined variational principle 
for the gravitational variables $\sigma$ and $\tensor ^4<A>$ by adding a 
second boundary term, now at $\Delta$ \cite{abck}.  To make the variational 
principle well-defined for the Maxwell connection $\emA$, one has to gauge-fix 
$\tensor ^4<\emA>$ partially at $\Delta$.  As in the definition of $\kappa$, 
our choice is designed to accomodate the standard static connection $\tensor 
^4<\emA>$ in the Reissner-Nordstr\"om solutions: We set $\tensor ^4<\emA> 
\cdot \ell = Q/r_\Delta$ on the horizon.  Also, in what follows, we will set 
the magnetic charge $P$ to zero for simplicity.

To pass to the Hamiltonian framework, one performs a Legendre transform.  Let 
us foliate the space-time $\mathcal{M}$ by partial Cauchy surfaces $M$ with 
inner boundaries $S_\Delta$ and outer boundaries at spatial infinity.  The 
phase space consists of quadruplets $(\tensor <A_a^AB>, \tensor 
<\Sigma_ab^AB>, \emA_a, \emE_{ab})$ satisfying appropriate boundary 
conditions, where $\tensor <A_a_A^B>$ and $\emA_a$ are the pull-backs to $M$ 
of $\tensor^4<A_a_A^B>$ and $\tensor^4<\emA_a>$, $\tensor <\Sigma_ab^AB>$ is 
the pull-back to $M$ of the 2-form $2\tensor <\sigma_[a^AA'><\sigma_b]^B_A'>$, 
and $\emE_{ab}$ is the electric field 2-form of the Maxwell field.  The 
symplectic structure is given by
\begin{equation}
  \begin{eqtableau}{1}
    \Omega(\delta_1, \delta_2) &= \frac{-i}{8\pi G} \left\{ \int_{M} \Tr 
      \left[ \delta_1 \Sigma \wedge \delta_2 A - 
             \delta_2 \Sigma \wedge \delta_1 A \right] +     \nonumber
      4 \oint_{S_\Delta} \Tr \left[ 
        \delta_1 (r_\Delta A) \wedge \delta_2 (r_\Delta A) \right] \right\} \\
    &\qquad\qquad - \int_{M} \left( \delta_2 \emA \wedge \delta_1 \emE - 
      \delta_1 \emA \wedge \delta_2 \emE \right),
  \end{eqtableau}
\end{equation}
where $\delta_1$ and $\delta_2$ represent any two tangent vectors to the phase 
space.

Fix a time-like vector field $t^a$, transverse to the leaves $M$ which equals 
$\ell^a$ on $\Delta$ and tends to an unit time translation orthogonal to $M$ 
at spatial infinity.  The corresponding Hamiltonian $H_t$ is given by
\begin{equation}\label{Ham}
  H_t = \int_\Sigma \mbox{constraints} + \underbrace{ 
  \lim_{r\rightarrow\infty}\, \frac{1}{4\pi}
     \oint_{S_r} \left( -\frac{r}{G} \Psi_2  \right)
		 \tensor ^2<\epsilon>}_{E_\mathrm{ADM}} - 
   \underbrace{\frac{1}{4\pi}\oint_{S_\Delta} \left( 
     -\frac{r_\Delta}{G} \Psi_2 + \frac{Q}{r_\Delta} \phi_1 \right) 
     \tensor ^2<\epsilon>}_{M_{\Delta}},
\end{equation}
where $\phi_1$ is a Newman-Penrose component of the Maxwell field tensor.  
Note that the surface term at infinity is precisely the ADM energy.  Indeed, 
in all physical theories, energy is the on-shell value of the generator of the 
appropriate time-translation.  For example, in space-times with several 
asymptotic regions, energy in any one `sector' is the on-shell value of the 
Hamiltonian which generates an unit time-translation in that region and no 
evolution in the others.  Hence, it is natural to interpret the surface term 
at $S_\Delta$ in (\ref{Ham}) as the energy associated with the isolated 
horizon corresponding to the time translation $\ell^a$.  Since $\ell^a$ 
defines the rest frame of the isolated black hole, this energy is the mass 
$M_\Delta$ of the isolated horizon.  The surface term can be re-expressed 
using the above definition of surface gravity as
\begin{equation}
  M_{\Delta} = \frac{1}{4\pi G} \kappa\area + \Phi Q,
\end{equation}
where $\Phi = Q/r_\Delta$ is the electric potential of the horizon.  Thus, 
when $\kappa$ and $M_{\Delta}$ are defined appropriately, Smarr's formula 
\cite{smarr} for the ADM mass of the Reissner-Nordstr\"om black holes extends 
to all isolated horizons.

It follows immediately from the above remarks that the Hamiltonian $H_t$ 
vanishes identically in Reissner-Nordstr\"om space-times.  (This is actually 
to be expected from rather general considerations in symplectic geometry.)  
Moreover, we can evaluate $H_t$ in any space-time where $\Delta$ extends to 
future timelike infinity as in figure \ref{exam}a.  Hamilton's equations of 
motion read
\begin{equation}\label{HamEq}
  \delta H_t = \Omega (\delta, X_{H_t}).
\end{equation}
where $X_{H_t}$ is the Hamiltonian vector field of $H_t$.  One can show the 
surface term on the right side of this equation vanishes due to boundary 
conditions, leaving only a bulk contribution.  Furthermore, the bulk term can 
be shown to equal the $\delta$-variation of an integral over future null 
infinity $\scri^+$, provided the fields in question have suitable fall-off 
\cite{abf2}.  The value $\ERad$ of this integral is precisely the total flux 
of energy through $\scri^+$ \cite{as}.  Equivalently, $\ERad$ is the total 
energy contained in radiative modes of the gravitational and electromagnetic 
fields \cite{as}.  It follows from (\ref{HamEq}), together with the vanishing 
of both $H_t$ and $\ERad$ on stationary solutions, that $H_t = \ERad$ whenever 
the constraints are satisfied and the above fall-off conditions hold.  
Substituting this result in (\ref{Ham}) gives
\begin{equation}
  M_{\Delta} = E_\mathrm{ADM} - \ERad
\end{equation}
on shell; \textit{the black hole mass is the future limit of the Bondi mass}.  
Physically, this is exactly what one would expect to find in a space-time 
containing both a black hole and radiation.  To our knowledge, $M_{\Delta}$ 
does not agree with any of the proposed quasi-local mass expressions in the 
charged case.  Rather, $M_{\Delta}$ is `the mass of the black hole together 
with its static hair;' it includes the energy associated with static fields 
emanating from $\Delta$ but \textit{not} contributions due to radiative 
excitations outside $\Delta$.

We are now in a position to establish the analogue of the first law of black 
hole mechanics for general isolated horizons.  In our framework, it is natural 
to regard $r_\Delta$ and $Q$ as the independent variables and express other 
physical quantities associated with the isolated horizons in terms of them:
\begin{equation}
  \begin{eqtableau}[.1cm]{2}
    M_\Delta &= \frac{r_\Delta}{2} \left(1 + \frac{Q^2}{r_\Delta^2} \right) 
      &\hspace{1cm}  \area &= 4\pi r_\Delta^2 \\
    \kappa &= \frac{1}{2r_\Delta} \left( 1 - \frac{Q^2}{r_\Delta^2} \right) 
      &\hspace{1cm}  \Phi &= \frac{Q}{r_\Delta}
  \end{eqtableau}
\end{equation}
One can simply vary this expression of the black hole mass to find
\begin{equation}
  \delta M_{\Delta} = \frac{1}{8\pi G} \kappa\delta\area + \Phi \delta Q.
\end{equation}
This is the first law of black hole mechanics, generalized to isolated 
horizons.

\topic{Discussion}

In summary, we introduced boundary conditions to define the notion of a 
non-rotating isolated horizon.  A striking consequence of these conditions is 
that, even though the exterior admits an infinite number of radiative degrees 
of freedom, one can nonetheless associate with the horizon extrinsic 
parameters $\kappa$ and $M_{\Delta}$ which have several physically desirable 
properties.  In particular, the obvious generalizations of the zeroth and 
first laws of black hole mechanics hold.  Furthermore, the boundary conditions 
are also well-suited for obtaining Lagrangian and Hamiltonian frameworks.  
This feature enables one to quantize the system non-perturbatively and provide 
a statistical mechanical derivation of entropy \cite{abck}.

We conclude with a few remarks:

\begin{list}{\arabic{enumi})}{\usecounter{enumi}
  \setlength{\labelwidth}{\parindent}
  \setlength{\itemindent}{\parindent}
  \setlength{\listparindent}{\parindent}
  \setlength{\leftmargin}{0pt}  \setlength{\rightmargin}{0pt}}

\item Non-trivial examples of isolated horizons can be obtained as follows.  
First, consider the gravitational collapse of a spherically symmetric star 
(figure \ref{exam}a without radiation) and regard the horizon $\Delta$ and 
future null-infinity $\scri^+$ as `initial-value surfaces' for a new 
space-time.  Specify radiative data which is zero on $\Delta$ and non-zero on 
$\scri^+$ and evolve it back to obtain a space-time $\mathcal{M}$ admitting 
both radiation and an isolated horizon.  For a second example, begin with a 
Schwarzschild-Kruskal space-time and change the initial data outside $r=R>2M$ 
on a Cauchy surface to include radiation.  Evolve this data to obtain a 
space-time $\mathcal{M}$ containing an isolated horizon which persists until 
radiation intersects the event horizon of the original space-time.  With our 
next example, we see there can be isolated horizons which are not parts of an 
event horizon.  Consider the surface $\Delta_1$ in figure \ref{shell}.  It 
seems physically unreasonable to exclude $\Delta_1$, by fiat, from 
\textit{all} thermodynamical considerations and focus only on the event 
horizon $\Delta_2$, especially if the second collapse occurs a very long time 
after the first.  Indeed all our results apply to $\Delta_1$ as well.  
Therefore, the notion of an isolated horizon seems better suited to 
equilibrium thermodynamics.  Finally, cosmological horizons with 
thermodynamical properties, such as those in de Sitter space-times, are 
encompassed by our treatment even though the space-time does not contain a 
black-hole in the usual sense.

\begin{figure}
  \begin{center}
    \includegraphics[height=4cm]{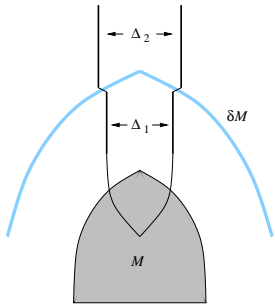} 
    \caption{A spherical star of mass $M$ undergoes collapse. Later, a 
      spherical shell of mass $\delta{M}$ falls into the resulting black 
      hole. While $\Delta_1$ and $\Delta_2$ are both isolated horizons,
			only $\Delta_2$ is part of the event horizon.}\label{shell}
  \end{center}
\end{figure}

\item The calculations of this paper have been performed using the connection 
dynamics framework \cite{blue} for general relativity.  Although the boundary 
conditions take a particularly compact form in terms of these variables, we 
fully expect the \textit{classical} framework can be formulated in terms of 
geometrodynamical variables.  For passage to non-perturbative quantization 
\cite{abck}, however, the connection variables seem essential.

\item Our framework is closely related to an interesting body of ideas 
developed by Hayward \cite{sh}.  Specifically, the isolated horizons 
introduced here are special cases of Hayward's trapping horizons.  The 
additional conditions imposed here were essential to our construction of 
Lagrangian and Hamiltonian frameworks, which in turn made it possible to carry 
out a non-perturbative quantization \cite{abck}.  Our strategies for defining 
$\kappa$ and $M_{\Delta}$ are also different from those introduced in 
\cite{sh}.  In particular, our definition yields the standard surface gravity 
for Reissner-Nordstr\"om black-holes, while the trapping gravity $\kappa$ of 
\cite{sh} does not.

\item There are at least three directions in which our framework should be 
generalized to encompass other interesting situations.  First, of course, 
rotating horizons should be allowed.  This would entail constructing local 
expressions for the angular momentum and rotational velocity of an isolated 
horizon.  At the present time, this appears to be mostly a technical matter as 
it requires one to weaken only the last boundary condition (on $n^a$) above.  
Second, Wald and collaborators have pointed out that the zeroth and first laws 
are largely theory independent in the stationary case \cite{wald}.  It would 
be interesting to attempt to extend that analysis to isolated horizons.  
Finally, one can envisage genuinely dynamical processes.  An extension of the 
framework along these lines would require non-trivial modifications but may be 
very useful, e.g., in numerical studies of black hole collisions.

\end{list}

\topic{Acknowledgments}

We would like to thank participants at the Third Mexican School on Gravitation 
and Mathematical Physics at Mazatl\'an for numerous comments.  This work was 
supported by the NSF grants PHY95-14240 and INT97-22514 and by the Eberly 
research funds of the Pennsylvania State University.

\end{document}